\pdfoutput=1
\documentclass[aps,prl,twocolumn,superscriptaddress,a4paper,
    floatfix]{revtex4}
\usepackage[latin1]{inputenc}
\usepackage{graphicx}
\usepackage{amsmath,amssymb}
\usepackage{hyperref}
\usepackage{datetime}

\begin{document}

\title{Observation of Quantized Conductance in Neutral Matter}

\author{Sebastian Krinner}
\affiliation{Department of Physics, ETH Zurich, 8093 Zurich, Switzerland}
\author{David Stadler}
\affiliation{Department of Physics, ETH Zurich, 8093 Zurich, Switzerland}
\author{Dominik Husmann}
\affiliation{Department of Physics, ETH Zurich, 8093 Zurich, Switzerland}
\author{Jean-Philippe Brantut}
\email{brantutj@phys.ethz.ch}
\affiliation{Department of Physics, ETH Zurich, 8093 Zurich, Switzerland}
\author{Tilman Esslinger}
\affiliation{Department of Physics, ETH Zurich, 8093 Zurich, Switzerland}

\date{\pdfdate}

\maketitle
{\bf
In transport experiments the quantum nature of matter becomes directly evident when changes in conductance occur only in discrete steps \cite{imry_quantCond}, with a size determined solely by Planck's constant $h$. The observations of quantized steps in the electric conductance \cite{van_wees_quantized_1988, wharam_one-dimensional_1988} have provided important insights into the physics of mesoscopic systems \cite{imry_introduction_2002} and allowed for the development of quantum electronic devices \cite{ihn_nanostructures}. Even though quantized conductance should not rely on the presence of electric charges, it has never been observed for neutral, massive particles \cite{sato_feasibility_2005}. In its most fundamental form, the phenomenon requires a quantum degenerate Fermi gas, a ballistic and adiabatic transport channel, and a constriction with dimensions comparable to the Fermi wavelength. 
Here we report on the observation of quantized conductance in the transport of neutral atoms. We employ high resolution lithography to shape light potentials that realize either a quantum point contact or a quantum wire for atoms. These constrictions are imprinted on a quasi two-dimensional ballistic channel connecting two adjustable reservoirs of quantum degenerate fermionic lithium atoms \cite{brantut_conduction_2012}. By tuning either a gate potential or the transverse confinement of the constrictions, we observe distinct plateaus in the conductance for atoms. The conductance in the first plateau is found to be equal to $1/h$, the universal conductance quantum. For low gate potentials we find good agreement between the experimental data and the Landauer formula, with all parameters determined a priori. Our experiment constitutes the cold atom version of a mesoscopic device and can be readily extended to more complex geometries and interacting quantum gases.
}

As pointed out by Landauer in 1957, conductance is the transmission of carriers from one terminal to another \cite{landauer_spatial_1957, buttiker_generalized_1985}. If the carriers move adiabatically through the channel connecting the terminals, each of its transverse modes contributes with $1/h$ to the conductance
\begin{equation}\label{equ}
G = \frac{1}{h}\sum_n f(E_n-\mu),
\end{equation}
where $f$ is the Fermi-Dirac distribution, $E_n$ the energy of the $n$-th transverse mode, and $\mu$ the chemical potential in the two terminals at equilibrium. When the temperature is sufficiently low compared to the transverse energy level spacing, the contribution of individual modes can be isolated in a transport measurement, leading to quantized plateaus in the conductance.

\begin{figure*}[htb]
    \includegraphics{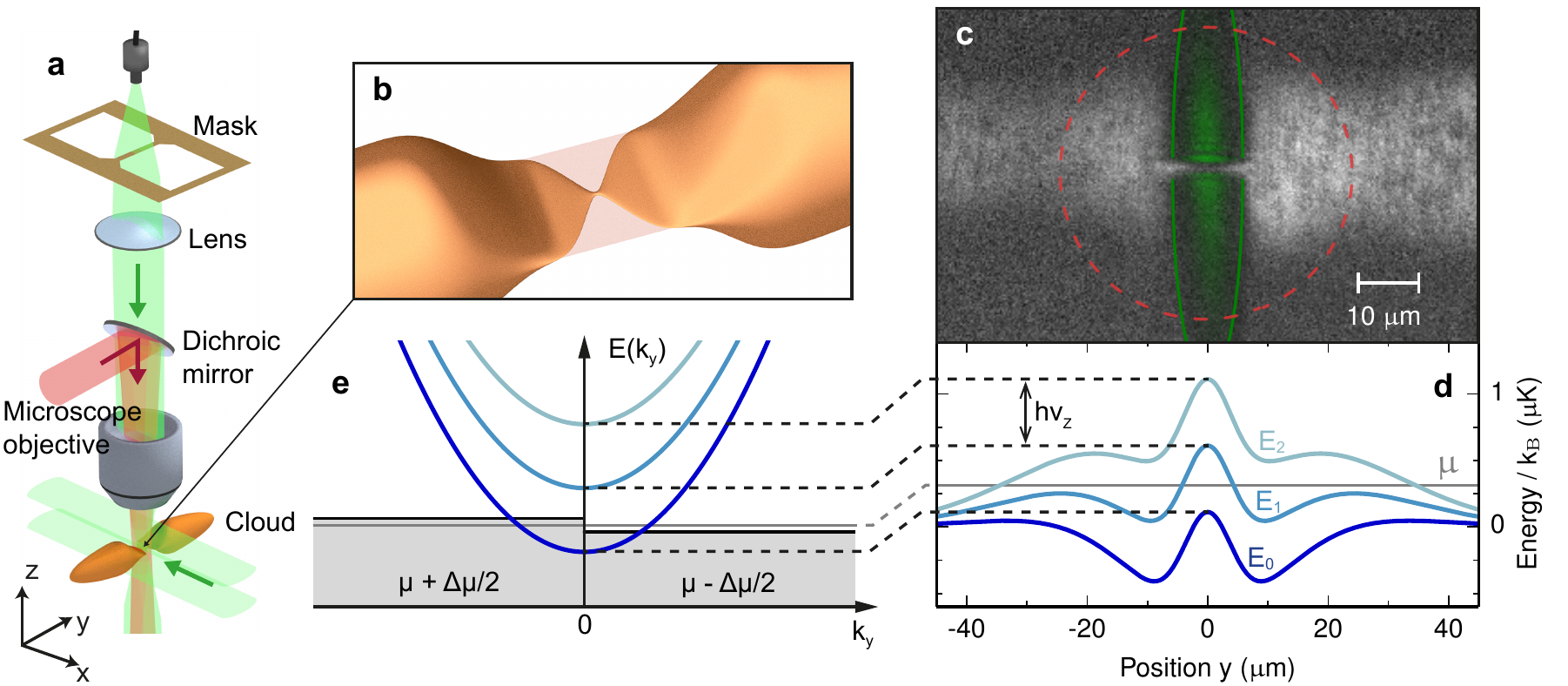}
    \caption{{\bf An atomic QPC.} a, Lithographic imprinting of the QPC (vertical green beam). An achromatic lens and a high numerical aperture microscope objective are used to demagnify the QPC structure onto the 2D channel region in the atomic cloud. An attractive gate potential is created by a red detuned laser beam (red beam). It is combined with the green beam on a dichroic mirror and focused onto the center of the QPC. The $\rm{TEM}_{01}$-like laser mode creating the 2D channel is shown as a horizontal green beam propagating along the $x$ axis. b, Zoom into the channel region: reservoirs, 2D channel and QPC are smoothly connected to each other. 
    c, Imaged atomic density in the QPC for $\nu_z=4.6\,\rm{kHz}$, $\nu_x=29\,\rm{kHz}$ and $V_g=1.4\,\rm{\mu K}$ in grey scale. The overlaid elliptic green region with the horizontal cut is an image of the projected split gate structure. Green solid line and red dashed line are $1/\text{e}^2$ contours of the split gate structure and of the gate potential, respectively. 
    d, Effective potential along the transport axis, consisting of the transverse mode energy and the gate potential (see Methods). The energy levels corresponding to the three lowest transverse modes are labeled with $E_0$, $E_1$ and $E_2$. The depicted situation corresponds to the first plateau of the $\nu_z=10.4\,\rm{kHz}$ data of Fig. \ref{fig2}. e, Energy dispersion relation of the particles in the QPC. The parabolas are offset by the quantized transverse mode energies. Left and right reservoirs are represented by grey boxes, having chemical potentials $\mu\pm\Delta\mu/2$.}\label{fig1}
\end{figure*}

Whilst ubiquitous in electronics, a two-terminal setup for atomic gases has only recently been demonstrated. Here, neutral atom currents play the role of electric currents and they are driven by a chemical potential bias rather than an electric voltage, a situation corresponding to ideal charge screening. Until now, the conductance of quasi two-dimensional multimode channels was measured, probing ballistic, diffusive and superfluid regimes \cite{brantut_conduction_2012, stadler_observing_2012, Krinner:2013aa}. Yet, in order to identify the contribution of an individual mode to the transport, a tightly confining channel geometry is required in which the energy separation between individual transverse modes is large compared to the temperature \cite{thywissen_quantum_1999}. 
This requirement is also encountered in quasi one-dimensional quantum gas experiments without reservoirs \cite{gorlitz_realization_2001, moritz_exciting_2003, paredes_tonks-girardeau_2004, kinoshita_observation_2004, bouchoule_atomchips2010, serwane_deterministic_2011}.
The basis of the experiment is our transport setup \cite{brantut_conduction_2012}. In brief, a weakly interacting gas of $N=7.5(3)\times 10^4$ fermionic $^6\rm{Li}$ atoms is prepared in a cigar-shaped trap at a temperature of $T=42(8)\,\rm{nK} = 0.11(2)T_F$, where $T_F=385(12)\,\rm{nK}$ denotes the Fermi temperature. The elongated trap is then split into two reservoirs connected by a two-dimensional channel using the repulsive potential of a $\rm{TEM}_{01}$-like mode of a laser operating at 532 nm, see Fig. \ref{fig1}a.

A quantum point contact (QPC) \cite{houten_quantum_1996} is created by lithographically projecting a split gate structure onto the two-dimensional channel, see Fig. \ref{fig1}a and b. To do so the negative of a slit with a width of $12\,\rm{\mu m}$ is printed on a binary mask and illuminated with a laser beam at 532\,nm. A projection system, consisting of an achromatic lens and a high numerical aperture microscope objective, demagnifies the object by a factor 11. The width of the resulting QPC in the channel region is measured to be $1.5\,(3)\,\rm{\mu m}$ (FWHM) in $x$ direction, using a second identical microscope placed opposite to the first one \cite{zimmermann_high-resolution_2011}. Since this value is comparable to the Fermi wavelength of $2.2\,\rm{\mu m}$ in our system, a single mode regime should be accessible. 
The finite width of the diffraction limited point spread function of the projection system leads to a harmonically confining potential along the $x$ axis. The overall shape of the projected QPC potential is given by the Gaussian profile of the illuminating laser beam, whose waists are $5.6(3)\,\rm{\mu m}$ ($33.6(6)\,\rm{\mu m}$) along the $y$ ($x$) axis in the image plane.

The QPC is characterized by its trap frequencies along $z$ and $x$ at its center, $\nu_z$ and $\nu_x$, originating from the harmonic confinement of the $\rm{TEM}_{01}$-like laser beam and the lithographically imprinted constriction respectively. Typical values are $\nu_z = 10.0(4)\,\rm{kHz}$ and $\nu_x = 30(3)\,\rm{kHz}$. The three lowest modes are thus separated by $h \nu_z \simeq 0.5\,\rm{\mu K}$, which is much more than the temperature of the gas.
The zero-point energy in the QPC is $E_0=(h\nu_z+h\nu_x)/2 \simeq 1.0\,\rm{\mu K}$, which is larger than the chemical potential $\mu = 370(11)\,\rm{nK}$, imposed by the reservoirs. To successively populate the transverse modes of the QPC we use an additional laser beam creating an attractive gate potential $V_g$ at the position of the QPC, see Fig. \ref{fig1}c, d, e. The laser beam with a wavelength of 767\,nm has a waist of $25.0(6)\,\rm{\mu m}$ and is propagating along the $z$ axis.
To measure the conductance we prepare an initial particle number imbalance $\Delta N_0=(N_L-N_R)_0=0.40(2)N$ between the two reservoirs, with $N_L$ and $N_R$ denoting the particle number in the left and right reservoir respectively.
This leads to a chemical potential bias $\Delta\mu = 94(7)\,\text{nK}\ll h\nu_z$, driving a current $I=G\Delta\mu$ across the QPC. We access the conductance by measuring the relative particle number imbalance after $1.5\,\rm{s}$ of transport time, assuming linear response, and evaluating the compressibility of the reservoirs (see Methods). 

\begin{figure}[htb]
    \includegraphics[width=0.45\textwidth]{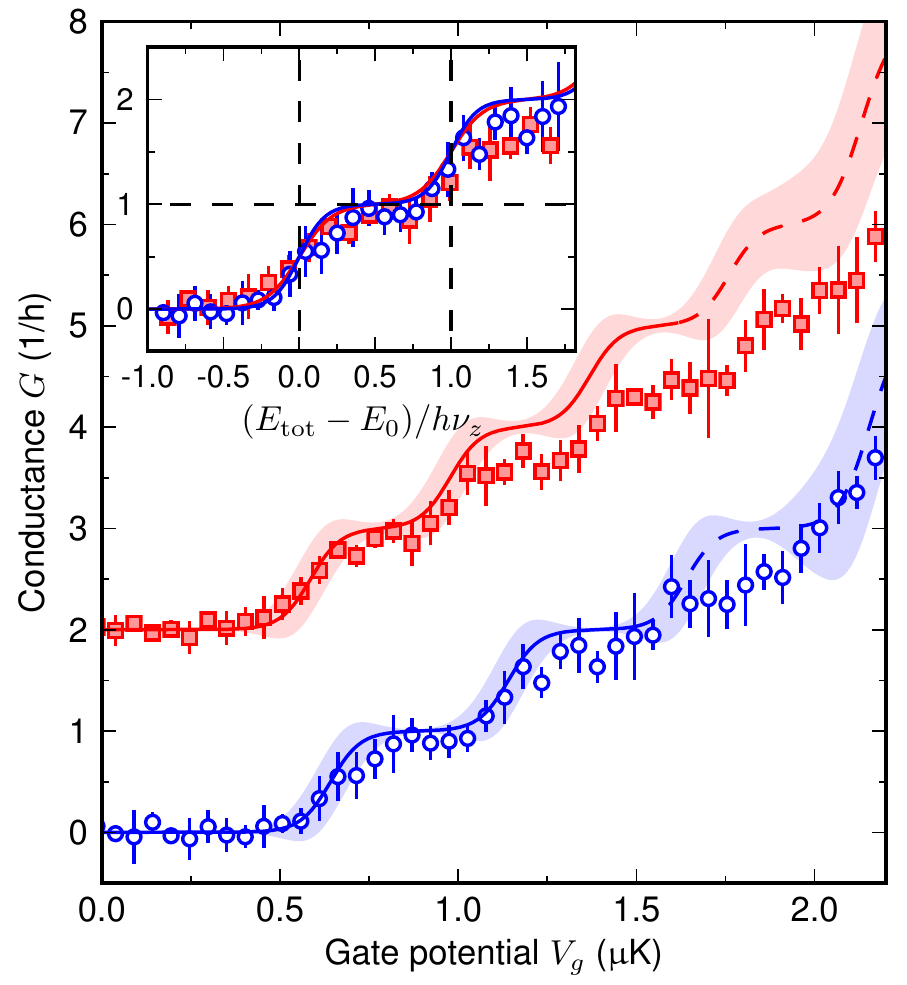}
    \caption{{\bf Conductance as a function of gate potential}. Open blue circles correspond to a vertical confinement of $\nu_z=10.4\,\rm{kHz}$. Filled red squares correspond to $\nu_z=8.2\,\rm{kHz}$ and are vertically shifted by two units for clarity. Each data point represents the mean of six measurements and error bars indicate one standard deviation. Solid lines are theoretical predictions based on the Landauer formula of conductance. The shaded regions reflect the uncertainties in the input parameters (see text). Dashed lines are continuations of the solid lines and correspond to a change in the effective potential (see Methods). 
    Inset: first conductance plateau as a function of reduced energy, showing universal scaling. Vertical dashed lines indicate the width of the first plateau, whereas the horizontal dashed line indicates the universal conductance value $1/h$.}
    \label{fig2}
\end{figure}

Fig. \ref{fig2} presents the measured conductance as a function of $V_g$ for two different vertical confinement frequencies, $\nu_z=10.4\,\rm{kHz}$ and $\nu_z=8.2\,\rm{kHz}$. The confinement along $x$ is set to $\nu_x=31.8\,\rm{kHz}$. Both curves start at zero conductance because the QPC is entirely closed at small gate potentials due to its zero-point energy. At the point where $V_g$ compensates the zero-point energy the conductance starts to rise and saturates at the universal value $1/h$ as soon as the ground state mode is tuned below the chemical potential of the reservoirs. 
Higher modes follow accordingly upon further increase of $V_g$. We clearly resolve the first two conductance plateaus in the case of $\nu_z=10.4\,\rm{kHz}$ (open blue circles), corresponding to the population of the $(n_x=0,n_z=0,1)$ modes, where $n_x$ and $n_z$ are the harmonic oscillator quantum numbers for the $x$ and $z$ direction respectively. 
For $\nu_z=8.2\,\rm{kHz}$ (filled red squares) the plateaus are narrower because the modes are more closely spaced in energy. In this case we resolve the first three modes, $(n_x=0,n_z=0,1,2)$ and even the onset of the $(0,3)$ mode is visible.

The inset of Fig. \ref{fig2} shows a zoom on the first conductance plateau. When substituting $V_g$ by the total energy $E_{\rm{tot}}=V_g+\mu$ of the particles minus $E_0$ and normalizing to $h\nu_z$, both data sets fall on top of each other as a consequence of universality, with the width of the plateau given by one unit of $h\nu_z$. The absolute accuracy of our conductance measurement is limited by the uncertainty in the compressibility of the reservoirs, which amounts to 11\% (see Methods). 

\begin{figure}[htb]
    \includegraphics[width=0.45\textwidth]{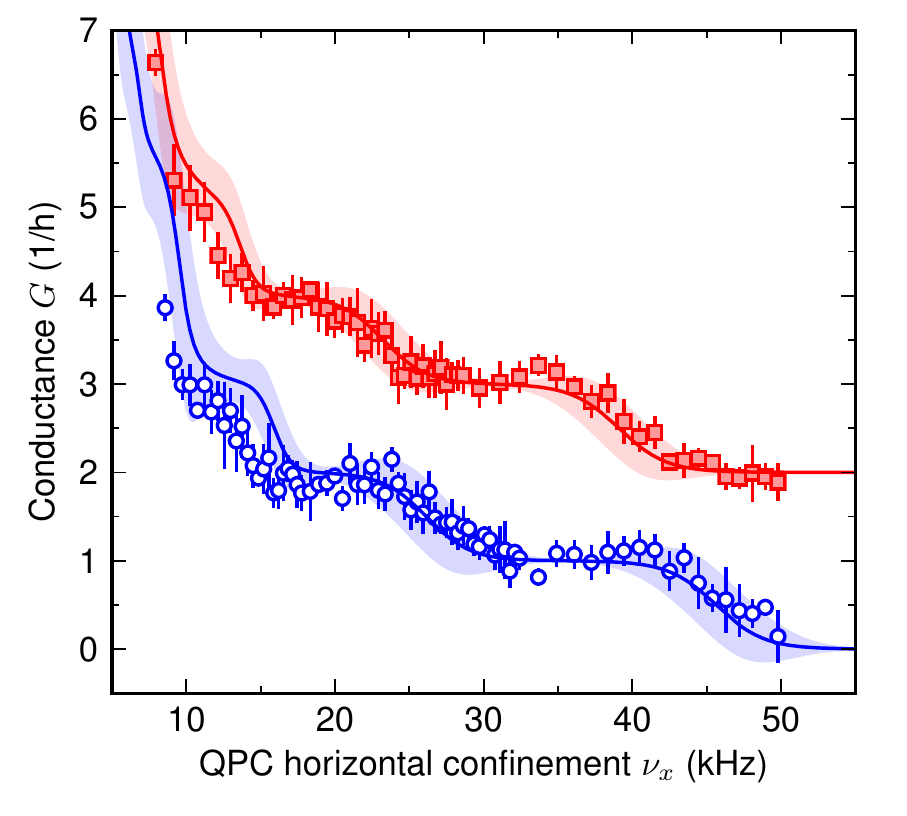}
    \caption{{\bf Conductance as a function of horizontal confinement.} Open blue circles correspond to a vertical confinement of $\nu_z=10.9\,\rm{kHz}$ and a gate potential of $V_g=1.0(1)\,\rm{\mu K}$. Filled red squares correspond to $\nu_z=9.2\,\rm{kHz}$ and $V_g=0.8(1)\,\rm{\mu K}$, and are vertically shifted by two units for clarity. Solid lines are theoretical predictions based on the Landauer formula of conductance. The shaded regions reflect the uncertainties in the input parameters (see text). Error bars are the same as in Fig. \ref{fig2}.}
    \label{fig3}
\end{figure}

The quantization of conductance is universal and should not depend on the control parameter. To demonstrate this, we next use the horizontal confinement $\nu_x$ of the QPC as a tuning parameter and keep the gate potential fixed. This is the counterpart of the celebrated measurement in solid state physics, where the split gate voltage is tuned to reveal quantized conductance \cite{van_wees_quantized_1988, wharam_one-dimensional_1988}. 
The blue data points in Fig. \ref{fig3} present this measurement for $\nu_z=10.9\,\rm{kHz}$ and $V_g=1.0(1)\,\rm{\mu K}$. We identify three increasingly wider plateaus centered at $\nu_x=12,\,20,\,35\,\rm{kHz}$. They correspond to the successive closing of the (1,0), (0,1) and (0,0) mode. 
The red data points in Fig. \ref{fig3} correspond to $\nu_z=9.2\,\rm{kHz}$ and  $V_g=0.8(1)\,\rm{\mu K}$ and show the same features as the blue data, with a reduced plateau width due to a smaller $\nu_z$.

We compare our data to theoretical predictions (solid lines in Fig. \ref{fig2} and \ref{fig3}) of the Landauer formula in the limit of entirely adiabatic and ballistic transport, Eqn. (\ref{equ}), without any fit parameter (see Methods). The input parameters $T$, $\mu$, $\Delta\mu$, $\nu_x$, $\nu_z$, $V_g$ are all independently measured quantities. Positions and widths of the plateaus are overall well predicted.
The conductances on the plateaus reach the universal values for moderate gate potentials. For larger gate potentials (see Fig. \ref{fig2}) the conductance is reduced with respect to the prediction. A possible reason is a small non-adiabaticity in the motion of the particles introduced by the gate potential, which accelerates the particles more and more towards the QPC as its strength is increased. 

The sharpness of the transition from one plateau to the next is set by the finite temperature. We checked numerically within the adiabatic approximation that broadening due to tunneling below the barrier or reflections above it \cite{glazman_reflectionless_1988} is much smaller than the relevant thermal broadening of the Fermi edge of $\sim$ $4\text{k}_\text{B} T$ \cite{ihn_nanostructures}. 
Further, we do not observe any non-linear effects \cite{kouwenhoven_nonlinear_1989} due to the applied finite bias $\Delta\mu$ because it is smaller than $\sim$ $4\text{k}_\text{B} T$, too.

\begin{figure}[htb]
    \includegraphics[width=0.45\textwidth]{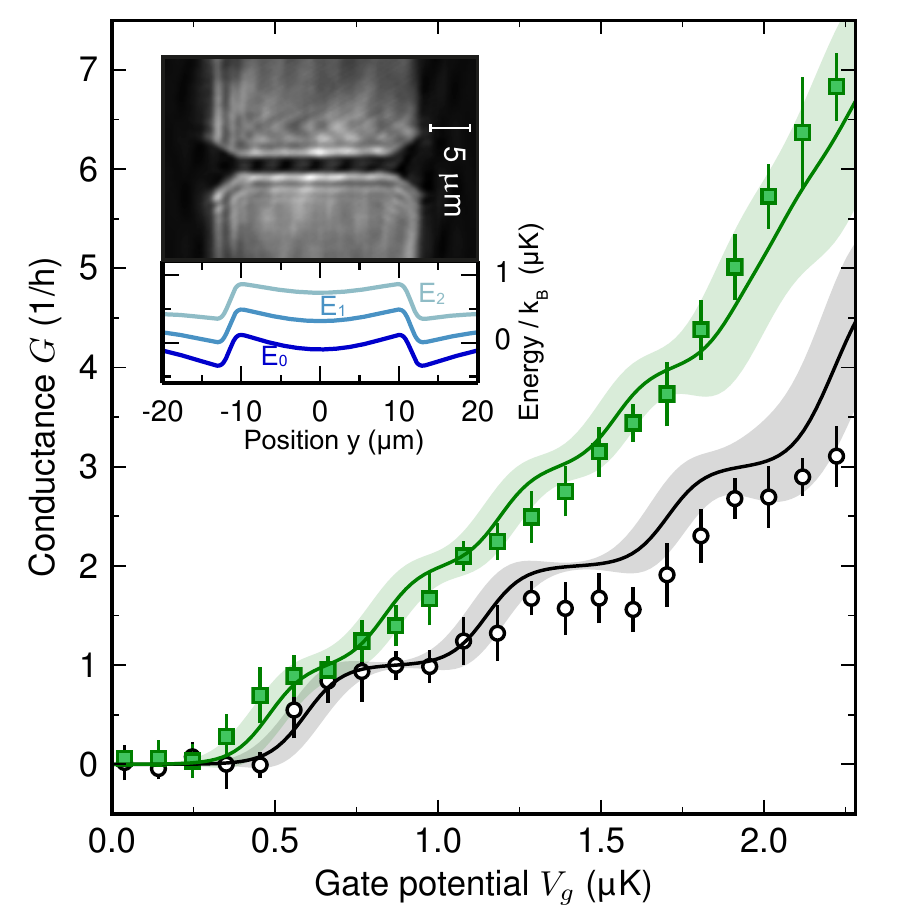}
    \caption{{\bf Quantum wire: conductance as a function of gate potential.} Open black circles correspond to a vertical confinement of $\nu_z=8.7\,\rm{kHz}$, and filled green squares to a lower confinement of $\nu_z=5.5\,\rm{kHz}$. For the stronger confinement clear conductance plateaus are observed. The weaker confinement illustrates the entrance of the first excited mode along the $x$ direction (see text). Solid lines are theoretical predictions based on the Landauer formula of conductance. Shaded regions and error bars are the same as in Fig. \ref{fig2}. The inset shows the lithographically imprinted wire as imaged by a second identical microscope objective. Below it, the corresponding effective potentials are drawn for the three lowest transverse modes, $E_0$, $E_1$, $E_2$, for the parameters at the first plateau of the $\nu_z=8.7\,\rm{kHz}$ data.}
    \label{fig4}
\end{figure}

The chemical potential or voltage in a ballistic constriction drops at its contacts \cite{ulreich_where_1998}, leading to universal conductance values independent of the length of the constriction. 
We demonstrate this by creating a long quantum wire (QW), see inset of Fig. \ref{fig4}. It has a length of $19.0(6)\,\rm{\mu m}$ and the same width of $1.5(3)\,\rm{\mu m}$ as the QPC.
The projected structure is spatially smoothed by the  low-pass filtering  of the projection system. The triangular-shaped openings further smoothen the transition from the QW to the 2D channel to avoid reflections at the openings \cite{szafer_theory_1989, yacoby_quantization_1990}.
The resulting effective potential, with the openings modeled by an error function, is shown in the inset of Fig. \ref{fig4}. It is highest at the entrance and exit of the QW, which is a consequence of the nearly uniform confinement in combination with the Gaussian envelope of the attractive gate beam.
The detection of single quanta of conductance is found to be less robust in this configuration, most likely because the points of highest potential are located at a slope of the gate beam. 
Nevertheless quantized conductance plateaus are observed for tight confinement ($\nu_x=24.6\,\rm{kHz}$, $\nu_z=8.7\,\rm{kHz}$, black circles in Fig. \ref{fig4}), using the gate potential as a tuning parameter. They are well reproduced by theory, up to a 10-20\% reduction in height for the second and third plateau. Possible interference effects, such as above-barrier resonances \cite{szafer_theory_1989} are not observed, likely because the finite temperature leads to a coherence length of the atoms of $7(1)\,\rm{\mu m}$, which is shorter than the length of the QW.

%
%

The green data in Fig. \ref{fig4} demonstrate how the plateaus disappear when $\nu_z$ is lowered to $5.5\,\rm{kHz}$.
The conductance starts to rise at a lower gate potential because of the reduced zero-point energy. Well defined plateaus are no longer visible due to the reduced ratio of $h\nu_z$ to temperature.
However we observe a change in slope of conductance by a factor of $\sim2$ at $V_g\simeq1.75\,\rm{\mu K}$, which is due to the entrance of the first excited state along the $x$ direction. More precisely, when $V_g$ is increased from $0.25$ to $1.75\,\rm{\mu K}$ the states $(n_x=0,n_z=0,1,2,3)$ enter, whereas above $V_g=1.75\,\rm{\mu K}$ the states $(n_x=0, n_z=4,5,...)$ and the states $(n_x=1,n_z=0,1,...)$ enter, giving rise to a change in slope by a factor of $2$. This feature is well covered by theory.

The demonstrated projection technique in combination with the ability to control all microscopic parameters a priori paves the way towards the quantum simulation of complex devices based on strongly correlated quantum gases. The detection of conductance at the level of single quanta provides access to the physics of topological edge states \cite{hasan_colloquium:_2010}, transport in the vicinity of a quantum phase transition \cite{sachdev_quantum_2011}, and universal conductance fluctuations \cite{lee_universal_1985}. 

We acknowledge fruitful discussions with G. Blatter, K. Ensslin, C. Glattli, T. Giamarchi, C. Grenier and M. Lebrat, and thank C. Chin, T. Ihn, Y. Imry, and W. Zwerger for their careful reading of the manuscript and discussions. We acknowledge financing from NCCR QSIT, the ERC Project SQMS, the FP7 project SIQS, and ETHZ. J.P.B. is supported by the Ambizione program of SNF.
\section{Methods summary}
A weakly interacting degenerate Fermi gas of $^6\rm{Li}$ atoms is prepared in a balanced spin mixture of the lowest and third lowest hyperfine states in a hybrid magnetic and optical dipole trap at a homogenous magnetic field of 552\,G, where the scattering length is -100$a_0$ with $a_0$ denoting Bohr's radius. To create a particle number imbalance between the two reservoirs we shift the trapping potential with respect to the QPC/QW along the $y$ axis during evaporative cooling. Shifting the trap back while blocking the transport with an additional repulsive gate laser beam \cite{brantut_thermoelectric_2013} results in a chemical potential bias. The transport process is started by removing the repulsive gate beam. After a transport time of 1.5\,s we switch it back on and measure the atom number in each reservoir. 
The temporal evolution of $\Delta N/N$ follows an exponential decay with a time constant $\tau=2G/C$, where $C$ is the compressibility of a single reservoir. 
We infer $\tau$ from the knowledge of the initial imbalance and the imbalance after 1.5\,s of transport time. Evaluating $C$ by assuming a non-interacting Fermi gas in a half-harmonic trap \cite{brantut_thermoelectric_2013} and taking into account the two-dimensional channel, directly yields $G$. All quantities in the text are stated for a single hyperfine state, implying that the conductance plateaus appear in multiples of 1/h and not 2/h. The transverse trapping frequencies of the QPC and QW are measured by parametric heating. The theory curves for the conductance rely on the adiabatic approximation \cite{glazman_reflectionless_1988, ihn_nanostructures}, allowing for a separation of longitudinal ($y$) and transverse ($x$, $z$) variables. In the resulting one-dimensional Schr\"{o}dinger equation the transverse energy $E_n(y)=E_{n_x,n_z}(y)$ and the gate potential $V_g(y)$ form the effective potential drawn in Fig. \ref{fig1}d and in the inset of Fig. \ref{fig4}. 

\section{Methods}
\subsection{Experimental Setup}
Our experiment uses the system described in \cite{brantut_conduction_2012,brantut_thermoelectric_2013}. In brief, a degenerate Fermi gas of $^6\rm{Li}$ atoms is produced by evaporative cooling of a balanced spin mixture of the lowest and third lowest hyperfine states in a hybrid magnetic and optical dipole trap at a homogenous magnetic field of 388\,G, using a magnetic field gradient \cite{hung_accelerating_2008}. The dipole trap operates at a wavelength of 1064\,nm, has a waist of $70\,\rm{\mu m}$ and a trap depth of $1.2\,\rm{\mu K}$. The experiments are performed in a homogeneous magnetic field of 552\,G, where the scattering length is $-100\,\rm{a_0}$, $\rm{a_0}$ denoting Bohr's radius. This allows for sufficiently fast thermalization of the reservoirs, while at the same time ensuring a ballistic transport channel. All quantities in the text are stated for a single hyperfine state, implying that the conductance plateaus appear in multiples of 1/h and not 2/h.

\subsection{Transport Sequence}
During the preparation of the degenerate gas, a magnetic field gradient of 0.2\,mT/m is applied along the $y$ axis in order to shift the trap with respect to the QPC. A repulsive elliptic gate beam focused onto the center of the QPC is used to separate the two reservoirs as in \cite{brantut_thermoelectric_2013}. The power of the laser beams creating the two-dimensional channel is ramped from zero to its final value within 200\,ms. The same ramp is applied to the laser beam creating the QPC. Then, evaporative cooling is enforced by using a magnetic field gradient along the $z$ axis \cite{hung_accelerating_2008}. This procedure results in a well defined particle number imbalance between the two reservoirs. 
Next, the dipole trap is adiabatically decompressed from a trap depth of $5.6\,\rm{\mu K}$ to a final depth of $1.2\,\rm{\mu K}$ within 200\,ms to further reduce the absolute temperature of the gas. During the same time interval the attractive gate potential is ramped from zero to $V_g$. Finally, the magnetic field gradient along $y$ is ramped to zero within 60\,ms, resulting in a well-defined chemical potential difference $\Delta\mu$ between the two reservoirs. We start the transport process by removing the repulsive gate beam. After a transport time of 1.5\,s we switch back on the repulsive gate beam to stop the transport process and we measure the atom number in both reservoirs via absorption imaging.  

\subsection{Conductance Evaluation}
From the equation for the current, $I=G\Delta\mu$, we obtain that the temporal evolution of the particle number imbalance between left and right reservoirs, $\Delta N=N_L-N_R$, is governed in linear response by the equation
\begin{equation}\label{diffEqu}
\frac{\text{d}}{\text{d}t}(\Delta N/N)=-\frac{G}{C_{\text{eff}}}(\Delta N/N),
\end{equation}
where $N=N_L+N_R$ is the total atom number and $C_{\text{eff}}=(1/C_L+1/C_R)^{-1}$ is an effective compressibility determined by the compressibilities of the single reservoirs at equilibrium. In agreement with this equation we observe an exponential decay of the relative particle number imbalance as a function of time, the time constant being $\tau=C_{\text{eff}}/G$. We evaluate $G$ by measuring $\tau$ and evaluating $C_{\text{eff}}$. The error made when substituting $\Delta\mu$ by $\Delta N/C_{\text{eff}}$ to obtain the linear response Eqn. \ref{diffEqu} is smaller than 5\% at the largest value of $\Delta N/N=0.4$.

$\tau$ is determined by measuring the relative particle number imbalance at $t=0$ and after a transport time of $t_{\text{tr}}=1.5s$. From the solution of Eqn. (\ref{diffEqu}) we obtain
\begin{equation}
\frac{1}{\tau}=\frac{1}{t_{\text{tr}}}\text{log}\left(\frac{\Delta N}{N}(t=0)\right)-\frac{1}{t_{\text{tr}}}\text{log}\left(\frac{\Delta N}{N}(t=t_{\text{tr}})\right),
\end{equation}
where a constant offset of -0.20(1) is substracted from both imbalances. This offset originates partly from the relative alignment of the gate beam and the lithographic system, and for the data of Fig. \ref{fig2} it varied linearly from -0.20 to -0.16 over the shown range.

The compressibilities $C_L=C_R=C$ of the identical reservoirs are calculated from the trap geometry, particle number and temperature, assuming a non-interacting Fermi gas. In brief, the trapping potential is harmonic along the $x$ and $z$ direction and half-harmonic along the $y$ direction \cite{brantut_thermoelectric_2013} with trapping frequencies of (194,\,23.5,\,157)\,Hz along the $x$, $y$ and $z$ direction. The effect of the repulsive potential of the $\text{TEM}_{01}$-like laser mode, creating the two-dimensional channel, is to shift the chemical potential by $17\%$ towards larger values with respect to the unperturbed cloud. The compressibility is almost not affected. The systematic uncertainty in $C_{\text{eff}}$ amounts to 11\%, which is due to the calibration error in the total particle number and an uncertainty in the oveall trapping potential.

\subsection{Trapping Frequencies of the QPC/QW}
The transverse trapping frequencies of the QPC and QW, $\nu_x$ and $\nu_z$, are measured by parametric heating in a dipole trap created by a laser beam with a waist of $8\,\rm{\mu m}$ and a wavelength of 767\,nm, propagating along the $z$ axis. The observed resonances have relative widths (FWHM) of $\delta\nu_z/\nu_z=0.05$ and $\delta\nu_x/\nu_x=0.30(5)$.
$\nu_z$ is found to depend weakly on $\nu_x$. This is because the darkness of the projected QPC structure decreases due to diffraction when moving out of focus along the $z$ axis, thus creating an additional confinement along $z$. We measure this contribution to be $\tilde{\nu}_z=0.16\,\nu_x$. Hence $\nu_z$ is given by $\nu_z=\sqrt{\nu_{z,0}^2+(0.16\,\nu_x)^2}$, where $\nu_{z,0}$ is the trapping frequency in the absence of the QPC. The values of $\nu_z$ stated in the text for the data of Fig. \ref{fig3} are evaluated for $\nu_x=31.8\,\rm{kHz}$ in order to be comparable to the values set for the data of Fig. \ref{fig2}. For the shown range of $\nu_x\in[10,50]\,\rm{kHz}$, $\nu_z$ varies by 12\% (16\%) around its mean value for the $\nu_z=10.9\,\rm{kHz}$ ($9.2\,\rm{kHz}$) data.

\subsection{Adiabatic Approximation and Theory Curves}
The computation of the conductance makes use of the adiabatic approximation \cite{glazman_reflectionless_1988, ihn_nanostructures}, allowing for a separation of longitudinal ($y$) and transverse ($x$, $z$) variables. It neglects scattering between different transverse modes and is justified if the confinement of the constriction varies smoothly along the transport direction. This is to a good approximation the case for both the QPC with its gaussian envelope and the QW with its triangular-shaped openings that are smoothed by the inherent low-pass filtering of the projection system. In the resulting one-dimensional Schr\"{o}dinger equation the transverse energy $E_n(y)=E_{n_x,n_z}(y)=h\nu_xf_x(y)(n_x+1/2)+h\nu_zf_z(y)(n_z+1/2)$, with $f_{x,z}(y)$ describing the spatial variation of the trapping frequencies, acts as an additional potential. Together with the gate potential $V_g(y)$ it forms the effective potential drawn in Fig. \ref{fig1}d and in the inset of Fig. \ref{fig4}. 
For the QPC we have $f_{x,z}(y)=\text{exp}(-y^2/w_{x,z}^2)$, with $w_x=5.6(3)\,\rm{\mu m}$ and $w_z=30(1)\,\rm{\mu m}$, whereas for the QW $f_x$ is constant along the wire, with its edges modeled by an error function.


The theory lines in Fig. \ref{fig2} are dashed above $V_g\sim 1.5\,\rm{\mu K}$ because at this point the maximum of the effective potential moves from the center to the sides of the QPC, which is not taken into account in the theory. This effect is not expected to explain the observed shift of the conductance below the universal values for large gate potentials. 

The conductance is calculated from Eqn. \ref{equ}, which we obtain from the two-terminal Landauer formula in the adiabatic regime \cite{ihn_nanostructures}
\begin{equation}
G = \frac{1}{h}\sum_n \int\limits_{-\infty}^\infty \text{d}E\,T_n(E)\,\left(-\frac{\partial f}{\partial E}(E-\mu)\right)
\end{equation}
by setting the transmission probability $T_n(E)=\Theta(E-E_n)$, with $\Theta(E-E_n)$ the Heaviside step function. This substitution corresponds to the semiclassical approximation, where the transmission probability for particles is one if their total energy is larger than their transverse energy and zero otherwise. It neglegts tunneling below the barrier and reflections above it \cite{glazman_reflectionless_1988}, which would lead to a broadening of $T_n(E)$. We checked numerically that for our geometry this broadening is much smaller than the thermal broadening of the Fermi-Dirac distribution and can thus be neglegted.

Taking into account the gate potential for the conductance evaluation, Eqn. \ref{equ} reads: 
\begin{equation}
G=\frac{1}{h}\sum_n f\left(\frac{E_n-V_g-\mu}{k_B T}\right).
\end{equation}
In the case of the quantum wire $V_g$ is multiplied by a factor $\text{exp}(-9.5^2/25^2)$ to account for the fact that the points of largest effective potential are located $9.5\,\rm{\mu m}$ from the center of gate potential and QW. The shaded error regions cover statistical and systematic errors of the input parameters and are determined using gaussian error propagation. The main contribution in Fig. \ref{fig2} and \ref{fig4} is the uncertainty in $\nu_x$, whereas in Fig. \ref{fig3} it is $V_g$.

\bibliographystyle{naturemag}

\begin{thebibliography}{10}
\expandafter\ifx\csname url\endcsname\relax
  \def\url#1{\texttt{#1}}\fi
\expandafter\ifx\csname urlprefix\endcsname\relax\def\urlprefix{URL }\fi
\providecommand{\bibinfo}[2]{#2}
\providecommand{\eprint}[2][]{\url{#2}}

\bibitem{imry_quantCond}
\bibinfo{author}{Imry, Y.}
\newblock \emph{\bibinfo{title}{Physics of mesoscopic systems, \textup{in}
  Directions in Condensed Matter, \textup{edited by G. Grinstein, G. Mazenko}}}
  (\bibinfo{publisher}{Singapore, World Scientific}, \bibinfo{year}{1986}).

\bibitem{van_wees_quantized_1988}
\bibinfo{author}{van Wees, B.~J.} \emph{et~al.}
\newblock \bibinfo{title}{Quantized conductance of point contacts in a
  two-dimensional electron gas}.
\newblock \emph{\bibinfo{journal}{Phys. Rev. Lett.}}
  \textbf{\bibinfo{volume}{60}}, \bibinfo{pages}{848--850}
  (\bibinfo{year}{1988}).

\bibitem{wharam_one-dimensional_1988}
\bibinfo{author}{Wharam, D.~A.} \emph{et~al.}
\newblock \bibinfo{title}{One-dimensional transport and the quantisation of the
  ballistic resistance}.
\newblock \emph{\bibinfo{journal}{J. Phys. C: Solid State Phys.}}
  \textbf{\bibinfo{volume}{21}}, \bibinfo{pages}{L209} (\bibinfo{year}{1988}).

\bibitem{imry_introduction_2002}
\bibinfo{author}{Imry, Y.}
\newblock \emph{\bibinfo{title}{Introduction to Mesoscopic Physics}}
  (\bibinfo{publisher}{Oxford University Press}, \bibinfo{year}{2002}).

\bibitem{ihn_nanostructures}
\bibinfo{author}{Ihn, T.}
\newblock \emph{\bibinfo{title}{Semiconductor Nanostructures}}
  (\bibinfo{publisher}{Oxford University Press}, \bibinfo{year}{2010}).

\bibitem{sato_feasibility_2005}
\bibinfo{author}{Sato, Y.}, \bibinfo{author}{Eom, B.-H.} \&
  \bibinfo{author}{Packard, R.}
\newblock \bibinfo{title}{On the feasibility of detecting quantized conductance
  in neutral matter}.
\newblock \emph{\bibinfo{journal}{J Low Temp Phys}}
  \textbf{\bibinfo{volume}{141}}, \bibinfo{pages}{99--109}
  (\bibinfo{year}{2005}).

\bibitem{brantut_conduction_2012}
\bibinfo{author}{Brantut, J.-P.}, \bibinfo{author}{Meineke, J.},
  \bibinfo{author}{Stadler, D.}, \bibinfo{author}{Krinner, S.} \&
  \bibinfo{author}{Esslinger, T.}
\newblock \bibinfo{title}{Conduction of ultracold fermions through a mesoscopic
  channel}.
\newblock \emph{\bibinfo{journal}{Science}} \textbf{\bibinfo{volume}{337}},
  \bibinfo{pages}{1069--1071} (\bibinfo{year}{2012}).

\bibitem{landauer_spatial_1957}
\bibinfo{author}{Landauer, R.}
\newblock \bibinfo{title}{Spatial variation of currents and fields due to
  localized scatterers in metallic conduction}.
\newblock \emph{\bibinfo{journal}{{IBM} Journal of Research and Development}}
  \textbf{\bibinfo{volume}{1}}, \bibinfo{pages}{223--231}
  (\bibinfo{year}{1957}).

\bibitem{buttiker_generalized_1985}
\bibinfo{author}{B\"{u}ttiker, M.}, \bibinfo{author}{Imry, Y.},
  \bibinfo{author}{Landauer, R.} \& \bibinfo{author}{Pinhas, S.}
\newblock \bibinfo{title}{Generalized many-channel conductance formula with
  application to small rings}.
\newblock \emph{\bibinfo{journal}{Phys. Rev. B}} \textbf{\bibinfo{volume}{31}},
  \bibinfo{pages}{6207--6215} (\bibinfo{year}{1985}).

\bibitem{stadler_observing_2012}
\bibinfo{author}{Stadler, D.}, \bibinfo{author}{Krinner, S.},
  \bibinfo{author}{Meineke, J.}, \bibinfo{author}{Brantut, J.-P.} \&
  \bibinfo{author}{Esslinger, T.}
\newblock \bibinfo{title}{Observing the drop of resistance in the flow of a
  superfluid fermi gas}.
\newblock \emph{\bibinfo{journal}{Nature}} \textbf{\bibinfo{volume}{491}},
  \bibinfo{pages}{736--739} (\bibinfo{year}{2012}).

\bibitem{Krinner:2013aa}
\bibinfo{author}{Krinner, S.}, \bibinfo{author}{Stadler, D.},
  \bibinfo{author}{Meineke, J.}, \bibinfo{author}{Brantut, J.-P.} \&
  \bibinfo{author}{Esslinger, T.}
\newblock \bibinfo{title}{Superfluidity with disorder in a thin film of quantum
  gas}.
\newblock \emph{\bibinfo{journal}{Phys. Rev. Lett.}}
  \textbf{\bibinfo{volume}{110}}, \bibinfo{pages}{100601}
  (\bibinfo{year}{2013}).

\bibitem{thywissen_quantum_1999}
\bibinfo{author}{Thywissen, J.~H.}, \bibinfo{author}{Westervelt, R.~M.} \&
  \bibinfo{author}{Prentiss, M.}
\newblock \bibinfo{title}{Quantum point contacts for neutral atoms}.
\newblock \emph{\bibinfo{journal}{Phys. Rev. Lett.}}
  \textbf{\bibinfo{volume}{83}}, \bibinfo{pages}{3762--3765}
  (\bibinfo{year}{1999}).

\bibitem{gorlitz_realization_2001}
\bibinfo{author}{G\"{o}rlitz, A.} \emph{et~al.}
\newblock \bibinfo{title}{Realization of bose-einstein condensates in lower
  dimensions}.
\newblock \emph{\bibinfo{journal}{Phys. Rev. Lett.}}
  \textbf{\bibinfo{volume}{87}}, \bibinfo{pages}{130402}
  (\bibinfo{year}{2001}).

\bibitem{moritz_exciting_2003}
\bibinfo{author}{Moritz, H.}, \bibinfo{author}{St\"{o}ferle, T.},
  \bibinfo{author}{K\"{o}hl, M.} \& \bibinfo{author}{Esslinger, T.}
\newblock \bibinfo{title}{Exciting collective oscillations in a trapped {1D}
  gas}.
\newblock \emph{\bibinfo{journal}{Phys. Rev. Lett.}}
  \textbf{\bibinfo{volume}{91}}, \bibinfo{pages}{250402}
  (\bibinfo{year}{2003}).

\bibitem{paredes_tonks-girardeau_2004}
\bibinfo{author}{Paredes, B.} \emph{et~al.}
\newblock \bibinfo{title}{Tonks-girardeau gas of ultracold atoms in an optical
  lattice}.
\newblock \emph{\bibinfo{journal}{Nature}} \textbf{\bibinfo{volume}{429}},
  \bibinfo{pages}{277--281} (\bibinfo{year}{2004}).

\bibitem{kinoshita_observation_2004}
\bibinfo{author}{Kinoshita, T.}, \bibinfo{author}{Wenger, T.} \&
  \bibinfo{author}{Weiss, D.~S.}
\newblock \bibinfo{title}{Observation of a one-dimensional tonks-girardeau
  gas}.
\newblock \emph{\bibinfo{journal}{Science}} \textbf{\bibinfo{volume}{305}},
  \bibinfo{pages}{1125--1128} (\bibinfo{year}{2004}).

\bibitem{bouchoule_atomchips2010}
\bibinfo{author}{Bouchoule, I.}, \bibinfo{author}{van Druten, N.} \&
  \bibinfo{author}{Westbrook, C.}
\newblock \emph{\bibinfo{title}{Atom Chips and One-Dimensional Bose Gases,
  \textup{in} Atom Chips, \textup{edited by J. Reichel, V. Vuletic}}}
  (\bibinfo{publisher}{John Wiley \& Sons}, \bibinfo{year}{2010}).

\bibitem{serwane_deterministic_2011}
\bibinfo{author}{Serwane, F.} \emph{et~al.}
\newblock \bibinfo{title}{Deterministic preparation of a tunable few-fermion
  system}.
\newblock \emph{\bibinfo{journal}{Science}} \textbf{\bibinfo{volume}{332}},
  \bibinfo{pages}{336 --338} (\bibinfo{year}{2011}).

\bibitem{houten_quantum_1996}
\bibinfo{author}{Houten, H.~v.} \& \bibinfo{author}{Beenakker, C.}
\newblock \bibinfo{title}{Quantum point contacts}.
\newblock \emph{\bibinfo{journal}{Physics Today}}
  \textbf{\bibinfo{volume}{49}}, \bibinfo{pages}{22--27}
  (\bibinfo{year}{1996}).

\bibitem{zimmermann_high-resolution_2011}
\bibinfo{author}{Zimmermann, B.}, \bibinfo{author}{M{\"u}ller, T.},
  \bibinfo{author}{Meineke, J.}, \bibinfo{author}{Esslinger, T.} \&
  \bibinfo{author}{Moritz, H.}
\newblock \bibinfo{title}{High-resolution imaging of ultracold fermions in
  microscopically tailored optical potentials}.
\newblock \emph{\bibinfo{journal}{New Journal of Physics}}
  \textbf{\bibinfo{volume}{13}}, \bibinfo{pages}{043007}
  (\bibinfo{year}{2011}).

\bibitem{glazman_reflectionless_1988}
\bibinfo{author}{Glazman, L.~I.}, \bibinfo{author}{Lesovik, G.~B.},
  \bibinfo{author}{{Khmel'Nitski\v{i}}, D.~E.} \& \bibinfo{author}{Shekhter,
  R.~I.}
\newblock \bibinfo{title}{Reflectionless quantum transport and fundamental
  ballistic-resistance steps in microscopic constrictions}.
\newblock \emph{\bibinfo{journal}{Soviet Journal of Experimental and
  Theoretical Physics Letters}} \textbf{\bibinfo{volume}{48}},
  \bibinfo{pages}{238} (\bibinfo{year}{1988}).

\bibitem{kouwenhoven_nonlinear_1989}
\bibinfo{author}{Kouwenhoven, L.~P.} \emph{et~al.}
\newblock \bibinfo{title}{Nonlinear conductance of quantum point contacts}.
\newblock \emph{\bibinfo{journal}{Phys. Rev. B}} \textbf{\bibinfo{volume}{39}},
  \bibinfo{pages}{8040--8043} (\bibinfo{year}{1989}).

\bibitem{ulreich_where_1998}
\bibinfo{author}{Ulreich, S.} \& \bibinfo{author}{Zwerger, W.}
\newblock \bibinfo{title}{Where is the potential drop in a quantum point
  contact?}
\newblock \emph{\bibinfo{journal}{Superlattices and Microstructures}}
  \textbf{\bibinfo{volume}{23}}, \bibinfo{pages}{719--730}
  (\bibinfo{year}{1998}).

\bibitem{szafer_theory_1989}
\bibinfo{author}{Szafer, A.} \& \bibinfo{author}{Stone, A.~D.}
\newblock \bibinfo{title}{Theory of quantum conduction through a constriction}.
\newblock \emph{\bibinfo{journal}{Phys. Rev. Lett.}}
  \textbf{\bibinfo{volume}{62}}, \bibinfo{pages}{300--303}
  (\bibinfo{year}{1989}).

\bibitem{yacoby_quantization_1990}
\bibinfo{author}{Yacoby, A.} \& \bibinfo{author}{Imry, Y.}
\newblock \bibinfo{title}{Quantization of the conductance of ballistic point
  contacts beyond the adiabatic approximation}.
\newblock \emph{\bibinfo{journal}{Phys. Rev. B}} \textbf{\bibinfo{volume}{41}},
  \bibinfo{pages}{5341--5350} (\bibinfo{year}{1990}).

\bibitem{hasan_colloquium:_2010}
\bibinfo{author}{Hasan, M.~Z.} \& \bibinfo{author}{Kane, C.~L.}
\newblock \bibinfo{title}{Colloquium: Topological insulators}.
\newblock \emph{\bibinfo{journal}{Rev. Mod. Phys.}}
  \textbf{\bibinfo{volume}{82}}, \bibinfo{pages}{3045--3067}
  (\bibinfo{year}{2010}).

\bibitem{sachdev_quantum_2011}
\bibinfo{author}{Sachdev, S.}
\newblock \emph{\bibinfo{title}{Quantum Phase Transitions}}
  (\bibinfo{publisher}{Cambridge University Press}, \bibinfo{year}{2011}).

\bibitem{lee_universal_1985}
\bibinfo{author}{Lee, P.~A.} \& \bibinfo{author}{Stone, A.~D.}
\newblock \bibinfo{title}{Universal conductance fluctuations in metals}.
\newblock \emph{\bibinfo{journal}{Phys. Rev. Lett.}}
  \textbf{\bibinfo{volume}{55}} (\bibinfo{year}{1985}).

\bibitem{brantut_thermoelectric_2013}
\bibinfo{author}{Brantut, J.-P.} \emph{et~al.}
\newblock \bibinfo{title}{A thermoelectric heat engine with ultracold atoms}.
\newblock \emph{\bibinfo{journal}{Science}} \textbf{\bibinfo{volume}{342}},
  \bibinfo{pages}{713--715} (\bibinfo{year}{2013}).

\bibitem{hung_accelerating_2008}
\bibinfo{author}{Hung, C.-L.}, \bibinfo{author}{Zhang, X.},
  \bibinfo{author}{Gemelke, N.} \& \bibinfo{author}{Chin, C.}
\newblock \bibinfo{title}{Accelerating evaporative cooling of atoms into
  bose-einstein condensation in optical traps}.
\newblock \emph{\bibinfo{journal}{Phys. Rev. A}} \textbf{\bibinfo{volume}{78}},
  \bibinfo{pages}{011604} (\bibinfo{year}{2008}).

\end{thebibliography}

\end{document}